# Anapoles and other magneto-electric multipoles in BaFe$_2$Se$_3$

S W Lovesey[1,2] and D D Khalyavin[1]

[1]ISIS Facility, STFC Oxfordshire OX11 0QX, UK

[2]Diamond Light Source Ltd, Oxfordshire OX11 0DE, UK

Abstract

We submit that the magnetic space-group C$_a$c (#9.41) is consistent with the established magnetic structure of BaFe$_2$Se$_3$, with magnetic dipole moments in a motif that uses two ladders [Caron J M et al 2011 *Phys. Rev.* B **84** 180409(R)]. The corresponding crystal class m1' allows axial and polar dipoles and forbids bulk ferromagnetism. The compound supports magneto-electric multipoles, including a magnetic charge (monopole) and an anapole (magnetic toroidal dipole) visible in the Bragg diffraction of x-rays and neutrons. Our comprehensive simulation of neutron Bragg diffraction by BaFe$_2$Se$_3$ exploits expressions of a general nature that can be of use with other magnetic materials. Electric toroidal moments are also allowed in BaFe$_2$Se$_3$. A discussion of our findings for resonant x-ray Bragg diffraction illustrates the benefit of a common platform for neutron and x-ray diffraction.

1. Introduction

Structural and magnetic properties on the iron chalcogenide BaFe$_2$Se$_3$ are of current interest for a number of reasons. The structure at 5 K, derived from neutron diffraction data, is orthorhombic Pnma, and compensating magnetic order develops below T$_N$ = 256 K [1]. Magnetic moments are parallel to the a-axis in a motif that uses two ladders, which appears to go hand-in-hand with local displacements of the ferrous ions ($^5$D). No superconductivity is observed down to 1.8 K [1]. Simulations of the electronic structure suggest that is ferrielectric, with a polarization that nearly cancels between ladders [2]. Although BaFe$_2$Se$_3$ merits study in its own right, because of its intriguing properties, similarities it appears to have with iron-based superconductors adds significant momentum to fully understanding it. A mechanism of superconductivity in iron pnictides and iron chalcogenides is not settled despite many experiments and many theoretical speculations [3].

The present study exploits a magnetic space-group for BaFe$_2$Se$_3$ derived from the established chemical and antiferromagnetic structure to predict properties that can be tested by neutron and resonant x-ray diffraction. Magneto-electric multipoles are emphasised in the study - best known of which are magnetic charges (monopoles) and anapoles - that have been neglected in any shape or form in literature on BaFe$_2$Se$_3$ hitherto, despite being firmly established in other compounds including multiferroics [4 - 10]. We demonstrate that anapoles, and higher-order magnetic multipoles, may contribute to the diffraction of neutrons by magnetic BaFe$_2$Se$_3$, and provide a comprehensive investigation of conditions required.

Multipoles that are parity-even, and time-odd, are present in the formulation of magnetic neutron diffraction published by Trammell [11] in 1957, although half a century elapsed before they were fully diagnosed as such [12]. A detailed simulation of an iridate perovskite (Sr$_2$IrO$_4$) illustrates the extent of information about valence electrons available from Bragg diffraction experiments when interpreted in terms of atomic multipoles constrained by all elements of symmetry in the appropriate magnetic space-group [13]. Recently, parity-odd and time-odd (magneto-electric) multipoles in neutron scattering have been discussed for CuO and a high-T$_c$ superconductor YBCO [14, 15].

Similar amounts of valuable information, about just those electrons which take part in bonding, is available from resonant x-ray Bragg and many successful experiments have been reported [5 - 8]. Formulations of the two types of experiments, neutron and resonant x-ray Bragg diffraction, share a structure factor for ions in a magnetic unit-cell that is a common platform. Discrete symmetries of multipoles and elements of symmetry from the magnetic space-group in the structure factor are one and the same for the two types of experiments, and differences in intensities of Bragg spots are caused by differences in properties of the radiations alone.

Our communication starts with information on the structural and magnetic properties of $BaFe_2Se_3$ from which we derive a magnetic space-group. Atomic multipoles in neutron scattering are reviewed in Section 3. The structure factor for $BaFe_2Se_3$ is presented and discussed in Section 4. Application of the structure factor to neutron diffraction by $BaFe_2Se_3$ occupies Sections 5 and 6. We pay attention in the main text to the simplest approximation that uses dipoles (multipoles with rank K = 1). If called for by an interpretation of diffraction data, a much more complete representation of the unit-cell structure factor including quadrupoles (K = 2) and octupoles (K = 3) can be readily derived from expressions that occupy an Appendix. The expressions in question are of potential value for the interpretation of data on other magnetic materials and in this sense they are thus universal. Section 7 is given over to a brief discussion of resonant x-ray Bragg diffraction and observation of magnetic charge. A discussion of our investigation appears in Section 8.

2. Structural and magnetic properties of $BaFe_2Se_3$

The parent structure is orthorhombic Pnma (#62) with cell dimensions a = 11.8834 Å, b = 5.4141 Å, c = 9.1409 Å [1]. The magnetic space-group $C_a c$ (#9.41) we have derived is polar [2] and belongs to the crystal class m1'. (Magnetic space-groups are specified in terms of the Belov-Neronova-Smirnova notations [16].) Basis vectors of the monoclinic cell are {(0, 2, 0), (2, 0, 0), (0, −1, −1)}with respect to the parent structure. Unique $b_m$-axis. Cell dimensions $a_m$ = 10.8282 Å, $b_m$ = 23.7668 Å, $c_m$ = 10.6240 Å and obtuse β = 120.638°. The chemical and magnetic structures of of $BaFe_2Se_3$ are depicted in Figure 1, in which the two ladders of dipole moments are made evident. The $C_a c$ magnetic space-group is associated with the four-dimensional physically irreducible representation $mR_1R_2$ (magnetic order wavevector 1/2, 1/2, 1/2) of the parent Pnma space group [17]. The magnetic order-parameter takes the $(\eta_1, 0, \eta_3, 0)$ direction in the representation space and yields a magnetically-induced electric polarization through $\{P_b (\eta_1^2 + \eta_3^2)\}$ and $\{P_c (\eta_1^2 − \eta_3^2)\}$ free-energy invariants, where $P_b$ and $P_c$ are components of the bulk electric polarization along the b-axis and c-axis of Pnma, respectively.

Estimates of structure parameters $(x_j, y_j, z_j)$ with j = 1 – 4 for $C_a c$ are listed in Table 1. They are estimates in so far as they are derived from the parent Pnma [1] that contains structural symmetry not present in $C_a c$. Actual structure parameters for $C_a c$ are four sets of *independent* quantities devoid of simple relations that exist between our tabulated estimates.

Physical properties are referred to orthogonal axes (x, y, z) that we elect to be $(\mathbf{a^*}_m, \mathbf{b}_m, \mathbf{c}_m)$ using vectors $\mathbf{a}_m = a_m (\sin\beta, 0, \cos\beta)$, $\mathbf{b}_m = b_m (0, 1, 0)$, $\mathbf{c}_m = c_m (0, 0, 1)$ and reciprocal vectors $\mathbf{a^*}_m = 2\pi b_m c_m (1, 0, 1)/v_o$, $\mathbf{b^*}_m = 2\pi a_m c_m \sin\beta (0, 1, 0)/v_o$, $\mathbf{c^*}_m = 2\pi a_m b_m (− \cos\beta, 0, \sin\beta)/v_o$ with cell volume $v_o = a_m b_m c_m \sin\beta$. A Bragg wavevector $\mathbf{k} = (h, k, l)_m$ with integer Miller indices is related to Miller indices $(H_o, K_o, L_o)$ for the orthorhombic parent structure by 2 $K_o$ = h, 2 $H_o$ = k and $K_o + L_o$ = − l. Thus the magnetic order wavevector (1/2, 1/2, 1/2) in the parent structure is (1, 1, −1)$_m$ in the monoclinic structure.

3. Atomic multipoles

Multipoles encapsulate various ground-state properties of valence electrons, and they are defined by us to possess discrete symmetries [12, 18]. A magnetic multipole is time-odd, and only multipoles of this type are visible in magnetic neutron scattering. The magnetic dipole moment $\langle \mathbf{L} + 2\mathbf{S} \rangle$ is time-odd and parity-even. (We use $\langle ... \rangle$ to denote the expectation value of the enclosed quantum-mechanical operators.) A multipole of rank K with the same discrete symmetries as the magnetic moment is denoted $\langle T^K_Q \rangle$, and the 2K + 1 projections obey $-K \leq Q \leq K$. Time-odd and parity-odd multipoles are called magneto-electric multipoles. Magneto-electric dipoles, also known as magnetic toroidal moments, include a spin anapole $\langle \mathbf{S} \times \mathbf{n} \rangle$, where $\mathbf{n}$ is the (local) electric dipole operator, and it was first studied by Zel'dovich in the course of investigating electromagnetic interactions that violate parity [19] (the magnetic field distribution of an anapole is reproduced by a solenoid deformed into a torus). Magneto-electric multipoles constructed from $\mathbf{S}$ and $\mathbf{n}$ are denoted here by $\langle H^K_Q \rangle$ [14]. Non-zero values of $\langle H^K_Q \rangle$ are manifestations of a state of magnetic charge. An actual magnetic charge, or magnetic monopole, $\langle H^0_0 \rangle \propto \langle \mathbf{S} \cdot \mathbf{n} \rangle$ is not visible in neutron scattering. (The label magnetic charge is justified by the observation that a "magnetic charge" inserted in Maxwell's equations, with symmetries of the electric and magnetic field unchanged, is both time-odd and parity-odd.) However, $\langle \mathbf{S} \cdot \mathbf{n} \rangle$ does contribute to the resonant diffraction of x-rays [9, 20], and estimates of $\langle \mathbf{S} \cdot \mathbf{n} \rangle$ in lithium orthophosphates have been published [10]. Orbital variables do not form magnetic charge, because $\mathbf{L}$ and $\mathbf{n}$ are orthogonal, $(\mathbf{L} \cdot \mathbf{n}) = (\mathbf{n} \cdot \mathbf{L}) = 0$, while an orbital anapole can be created from $\mathbf{\Omega} = [\mathbf{L} \times \mathbf{n} - \mathbf{n} \times \mathbf{L}]$. Magneto-electric multipoles associated with orbital variables alone are denoted by $\langle O^K_Q \rangle$, and the dipole $\langle O^1 \rangle$ contains $\langle \mathbf{\Omega} \rangle$ [14]. The panoply of magneto-electric multipoles reviewed by Dubovik and Tugushev [21] includes the Majorana fermion.

Magneto-electric multipoles are zero if the point group contains a centre of inversion symmetry. In the absence of this symmetry element expectation values $\langle H^K_Q \rangle$ and $\langle O^K_Q \rangle$ can be different from zero, because orbitals with angular momenta that differ by an odd integer are admixed. For a 3d ion the admixture may contain 3d(Fe) - 4p(Fe) in addition to charge transfer using 3d(Fe) and p-states of ligand ions, for example. Magnetic properties of a ferrous ion ($Fe^{2+}$, $3d^6$) are likely strongly influenced by interactions engaging the valence electron residing outside a shell that is half full. Model calculations that illustrate these feature of magneto-electric multipoles are found in references [18, 20, 22].

4. Structure factor for Bragg diffraction

Unit-cell structure factors for neutron and x-ray diffraction are constructed from,

$$\Psi^K_Q = \sum_\mathbf{d} \exp(i\mathbf{d} \cdot \mathbf{k}) \langle U^K_Q \rangle_\mathbf{d}. \qquad (4.1)$$

Sites in a unit cell labelled by vectors $\mathbf{d}$ are occupied by magnetic ferrous ions that possess a nominal magnetic moment = 4 $\mu_B$ in the high-spin configuration. The structure factor $\Psi^K_Q$ complies with all elements of symmetry in a magnetic space-group, because translation symmetry is used to relate environments of ions. In the magnetic space-group $C_ac$ electronic multipoles $\langle U^K_Q \rangle$ are not constrained by symmetry. Bulk properties of a material are described by $\Psi^K_Q$ evaluated for $\mathbf{k} = 0$ and they are prescribed by elements of symmetry in the crystal class. Additional physical properties of the material are allowed by the magnetic space-group and may contribute to Bragg spot intensities.

It is convenient in calculations of unit-cell structure factors to divide $\Psi^K_Q$ into its even and odd parts with respect to Q. We choose $\Psi^K_Q = A^K_Q + B^K_Q$ and $\Psi^K_{-Q} = A^K_Q - B^K_Q$.

Positions of the 32 ferrous ions in $C_ac$ are defined by four sets of structure parameters and our estimates are found in Table 1. We employ spatial phase factors $f_j = \exp(2\pi i(x_j h + z_j l))$ and $g_j = \exp(2\pi i y_j k)$, and go on to find a general result applicable to both neutron and x-ray diffraction,

$$\left.\begin{array}{l} A^K_Q \\ \\ B^K_Q \end{array}\right| = (1/2)[\langle U^K_Q\rangle \pm \langle U^K_{-Q}\rangle][1 + (-1)^{h+k}][1 + \sigma_\theta(-1)^k]\sum_j f_j[g_j \pm g_j^* \sigma_\theta \sigma_\pi (-1)^{1+K+Q}], \quad (4.2)$$

where the upper sign in $\pm$ belongs to $A^K_Q$ and the lower sign belongs to $B^K_Q$. In the expression $\sigma_\theta = \pm 1$ and $\sigma_\pi = \pm 1$ are signatures of time and parity for the multipole $\langle U^K_Q\rangle$. For magnetic neutron scattering $\sigma_\theta = -1$, while $\sigma_\pi = +1$ for $\langle T^K_Q\rangle$ and $\sigma_\pi = -1$ for $\langle H^K_Q\rangle$ and $\langle O^K_Q\rangle$. According to (4.2), diffraction from ferrous ions can occur provided (h + k) is even, and magnetic diffraction requires k odd, i.e., h and k must be odd for magnetic Bragg diffraction by $BaFe_2Se_3$.

Multipoles we consider are created from Hermitian operators and they satisfy $\langle U^K_{-Q}\rangle = (-1)^Q \langle U^K_Q\rangle^*$. We write a multipole in terms of its real and imaginary parts with the phase convention $\langle U^K_Q\rangle = \langle U^K_Q\rangle' + i\langle U^K_Q\rangle''$, leading to $A^K_Q \propto \langle U^K_Q\rangle'$ for Q even and $A^K_Q \propto i\langle U^K_Q\rangle''$ for Q odd, and vice versa for $B^K_Q$.

On using h = k = l = 0 and $\sigma_\theta = -1$ in (4.2) the electronic structure factor for bulk properties is zero, because the magnetic structure is fully compensated. Bulk axial and polar dipoles can be different from zero, however. For inspection of (4.2) with K = 1 and $\sigma_\theta = +1$ shows the structure factor can have components parallel to $\mathbf{a}^*_m$ and $\mathbf{c}_m$ using $\sigma_\pi = -1$, while with K = 1 and $\sigma_\theta = +1$ a component exists parallel to $\mathbf{b}_m$ using $\sigma_\pi = -1$. These structural properties, bulk polar and axial moments, are properties of the crystal class m1', as is the absence of bulk ferromagnetism.

Magnetic charge $\langle \mathbf{S} \cdot \mathbf{n}\rangle$ can be observed in resonant x-ray diffraction but not in dichroic signals [20]. The relevant quantity is the scalar structure factor $\Psi^0_0 = A^0_0$ with $\sigma_\theta = \sigma_\pi = -1$. From (4.2) we find $\Psi^0_0$ with $\sigma_\theta = \sigma_\pi = -1$ can different from zero for $BaFe_2Se_3$. Likewise, $\Psi^0_0$ can be different from zero for $\sigma_\theta = -1$ and $\sigma_\pi = +1$. The corresponding monopole can be represented by the scalar product of the magnetic dipole ($\sigma_\theta = -1$ and $\sigma_\pi = +1$) and an electric toroidal moment ($\sigma_\theta = +1$ and $\sigma_\pi = +1$). A a structural rotation affords a physical representation of an electric toroidal moment, i.e., the octahedral tilting in perovskites or any other polyhedral rotation. 5. A vector product of the two dipole operators, magnetic moment and electric toroidal moment, is equivalent to another "magnetic moment" but this dipole nor the corresponding monopole contribute in diffraction experiments, to the best of our knowledge.

The amplitude for magnetic neutron diffraction is $\langle\mathbf{Q}_\perp\rangle = [\boldsymbol{\kappa} \times (\langle\mathbf{Q}\rangle \times \boldsymbol{\kappa})]$ using a unit vector $\boldsymbol{\kappa} = \mathbf{k}/k$. The intermediate amplitude $\langle\mathbf{Q}\rangle$ contains $\langle\exp(i\mathbf{k}\cdot\mathbf{R})\mathbf{S}\rangle$ where $\mathbf{S}$ and $\mathbf{R}$ are electron spin and position operators. On treating $\mathbf{k}$ as a small quantity, $\langle\mathbf{Q}\rangle$ is proportional to $\langle\mathbf{S}\rangle$ at the initial level of an expansion. A moving charge creates a magnetic field so there is a contribution to $\langle\mathbf{Q}\rangle$ from the linear momentum of an electron, $\mathbf{p}$, in addition to its spin. Some fiddly algebra is required to produce a contribution proportional to orbital angular momentum, $\mathbf{L} = \mathbf{R} \times \mathbf{p}$, in the second level of an expansion of $\langle\exp(i\mathbf{k}\cdot\mathbf{R})(\boldsymbol{\kappa} \times \mathbf{p})/i\hbar k\rangle$. Combining the two results, we may write $\langle\mathbf{Q}\rangle \approx (1/2) f(k) \langle\mathbf{L} + 2\mathbf{S}\rangle$ in which the atomic form factor $f(k)$ is constructed from radial integrals with $f(0) = 1$. This small $\mathbf{k}$ limit for $\langle\mathbf{Q}\rangle$ was published by Schwinger in 1937 [23] and it is routinely exploited to determine motifs of magnetic dipoles, $\langle\mathbf{L} + 2\mathbf{S}\rangle$, in magnetic materials. Additional time-odd and parity-even multipoles $\langle T^K_Q \rangle$ in the diffraction amplitude include a quadrupole (K = 2) and and an octupole (K = 3) and associated form factors vanish in the forward direction of scattering. At the second level in an expansion in **k**, the spin contribution to $\langle \mathbf{Q} \rangle$ is parity-odd, namely, $ikR\langle(\boldsymbol{\kappa}\cdot\mathbf{n})\mathbf{S}\rangle$ in which $\mathbf{n} = \mathbf{R}/R$. It is straightforward to show that this contribution is equivalent to $(ikR/2)[\boldsymbol{\kappa} \times \langle \mathbf{S} \times \mathbf{n} \rangle + \langle \mathbf{S}(\boldsymbol{\kappa}\cdot\mathbf{n}) + (\boldsymbol{\kappa}\cdot\mathbf{S})\mathbf{n} - (2/3)\boldsymbol{\kappa}(\mathbf{S}\cdot\mathbf{n})\rangle]$. The spin anapole $\langle \mathbf{S} \times \mathbf{n} \rangle$ has been observed in several compounds using resonant x-ray diffraction [7, 8], and the quadrupole $\langle H^2_Q \rangle$ is derived from $\langle \mathbf{S}(\boldsymbol{\kappa}\cdot\mathbf{n}) + (\boldsymbol{\kappa}\cdot\mathbf{S})\mathbf{n} - (2/3)\boldsymbol{\kappa}(\mathbf{S}\cdot\mathbf{n})\rangle$. A magnetic charge $\langle H^0_0 \rangle \propto \langle \mathbf{S}\cdot\mathbf{n} \rangle$ is omitted from the development of $i\langle(\boldsymbol{\kappa}\cdot\mathbf{n})\mathbf{S}\rangle$, because it is multiplied by $\boldsymbol{\kappa}$ and does not therefore appear in the scattering amplitude $\langle \mathbf{Q}_\perp \rangle = \langle \mathbf{Q} \rangle - \boldsymbol{\kappa}(\boldsymbol{\kappa}\cdot\langle \mathbf{Q} \rangle)$. Use of an identity for the angular part of linear momentum reveals an orbital anapole in $\langle \mathbf{Q} \rangle$. The identity, or operator equivalent, in question is $\mathbf{p}_\omega \equiv (1/2R)[\boldsymbol{\Omega} - 2i\,\mathbf{n}]$, and $\langle [\boldsymbol{\Omega} - 2i\,\mathbf{n}] \rangle$ generates the dipole $\langle \mathbf{O}^1 \rangle$ [14].

Cartesian components of a dipole (K = 1) are constructed using,

$$\Psi^1_x = (\Psi^1_{-1} - \Psi^1_{+1})/\sqrt{2}, \quad \Psi^1_y = i(\Psi^1_{-1} + \Psi^1_{+1})/\sqrt{2}, \quad \Psi^1_z = \Psi^1_0. \tag{5.1}$$

The standard dipole-approximation for $\langle \mathbf{Q} \rangle$ is,

$$\langle \mathbf{Q} \rangle \approx (3/2)(\Psi^1_x, \Psi^1_y, \Psi^1_z), \tag{5.2}$$

in which $\Psi^1_p$ is evaluated with $\sigma_\pi = +1$ and $\langle \mathbf{U}^1 \rangle = \langle \mathbf{T}^1 \rangle$ [6, 12, 25].

Multipoles depend on the magnitude of the scattering wavevector through weighted integrals of the radial density of atomic wavefunctions. For parity-even multipoles $\langle T^K \rangle$ these so-called radial integrals are usually denoted $(j_n)$ where n is an even positive integer [24]. They are defined such that $(j_0) = 1$ for k = 0, while $(j_n) = 0$ for n ≥ 2. For the dipole approximation [25],

$$\langle \mathbf{T}^1 \rangle \approx (1/3)\{2\langle \mathbf{S} \rangle (j_0) + \langle \mathbf{L} \rangle [(j_0) + (j_2)]\}. \tag{5.3}$$

An even rank multipole is proportional to a radial integrals of the same order, i.e., $\langle T^K \rangle \propto (j_K)$ for K even [13], whereas an odd rank multipole is related to two radial integrals of order K ± 1 as in (5.3). Unexpected even-rank $\langle T^K \rangle$ arise from the spin and orbital contribution to $\langle \mathbf{Q} \rangle$ when the magnetic ground-state contains two or more manifolds of atomic states [25]. The potential importance of K-even multipoles has been illustrated in detailed simulations including K = 2 and 4 of diffraction by an iridate [13]. Turning to parity-odd multipoles, we see in $\langle \mathbf{Q} \rangle$ that both the spin anapole $\langle \mathbf{S} \times \mathbf{n} \rangle$ and quadrupole occur with a factor k, and thus radial integrals in $\langle H^1_Q \rangle$ and $\langle H^2_Q \rangle$ vanish in the limit k → 0. By contrast, the radial integral associated with the orbital anapole is unbounded as the wavevector tends to zero, because linear momentum in $\langle \mathbf{Q} \rangle$ is accompanied by a factor (1/k). In the case of $BaFe_2Se_3$ the smallest magnetic Bragg wavevector $\mathbf{k} = (1, 1, 0)_m$ has a magnitude k = 0.72 Å$^{-1}$.

Calculations of the amplitude for magnetic neutron diffraction that we report in an Appendix retain multipoles proportional to what are anticipated to be the largest radial integrals. The parity-even contribution we consider includes only dipoles, quadrupoles and octupoles. Parity-odd dipoles and quadrupoles are included in our chosen level of approximation. The full range of multipoles in neutron diffraction is catalogued in references [12, 14, 25, 26].

6. Neutron diffraction by BaFe$_2$Se$_3$

A unit-cell structure factor for neutron diffraction can be derived from (4.2). The result for $\langle \mathbf{Q}_\perp \rangle$ is a sum of multipoles that are time odd, and either parity-even or parity-odd. There are many multipoles in $\langle \mathbf{Q}_\perp \rangle$, because BaFe$_2$Se$_3$ possesses little symmetry and consequently few selection rules operate. Radial integrals discussed in the previous section are a major factor in controlling the magnitude of contributions, and high-order multipoles are often insignificant when low-order multipoles assume typical values. In this case, an expression for $\langle \mathbf{Q}_\perp \rangle$ using low-order multipoles will furnish a representation adequate for most purposes. To this end, an Appendix carries a result for $\langle \mathbf{Q}_\perp \rangle$ that is universal and correct to the level of parity-even octupoles and parity-odd quadrupoles. A result specific to BaFe$_2$Se$_3$ is obtained by inserting values from (4.2)

Let us explore intensities of Bragg spots (h, k, 0)$_m$ for BaFe$_2$Se$_3$ for which h and k are odd integers, $\kappa_x \propto h$, $\kappa_y \propto k$ and $\kappa_z = 0$. We retain only dipoles from the universal expression for the scattering amplitude reported in an Appendix, because these have the largest radial form factors at small Bragg wavevectors. The role of higher-order multipoles will be enhanced in the event that a ferrous ion has very small dipoles. However, the reported magnetic dipole moment is not small, with $\langle \mathbf{L} + 2\mathbf{S} \rangle \approx 2.8$ at 5 K [1]. We find an intensity,

$$| \langle \mathbf{Q}_\perp \rangle^{(+)} + \langle \mathbf{Q}_\perp \rangle^{(-)} |^2 \approx (3/2) | \sqrt{3}(\kappa_x A^1_1 - i\kappa_y B^1_1) + A^1_0 |^2$$
$$+ 3 | (\sqrt{3}/2) \mathcal{A}^1_0 + \kappa_x \mathcal{A}^1_1 - i\kappa_y \mathcal{B}^1_1 |^2. \qquad (6.1)$$

Here, $A^1_Q$ and $B^1_Q$ refer to parity-even dipoles (K = 1, $\sigma_\theta \sigma_\pi = -1$) and $\mathcal{A}^1_Q$ and $\mathcal{B}^1_Q$ refer to parity-odd dipoles (K = 1, $\sigma_\theta \sigma_\pi = +1$). Cartesian components of local ferrous dipoles $\langle \mathbf{T}^1 \rangle$ and $\langle \mathbf{O}^1 \rangle$ are purely real. From (4.2) we get $A^1_0 = 8 \langle T^1_z \rangle \aleph$, $A^1_1 = 4\sqrt{2} \langle T^1_y \rangle \Re$, $B^1_1 = -4\sqrt{2} \langle T^1_x \rangle \aleph$, $\mathcal{A}^1_0 = 8i \langle O^1_z \rangle \Re$, $\mathcal{A}^1_1 = -4i\sqrt{2} \langle O^1_y \rangle \aleph$ and $\mathcal{B}^1_1 = -4i\sqrt{2} \langle O^1_x \rangle \Re$, with complex numbers $\aleph = \sum_j f_j \, \text{Re}(g_j)$ and $\Re = \sum_j f_j \, \text{Im}(g_j)$ that depend on Miller indices h and k. These expressions yield a result for the intensity of the Bragg spot (h, k, 0)$_m$ in terms of the physically important dipoles,

$$| \langle \mathbf{Q}_\perp \rangle^{(+)} + \langle \mathbf{Q}_\perp \rangle^{(-)} |^2 \approx 96\{ | \sqrt{(3/2)} (\kappa_x \langle T^1_y \rangle \Re + i\kappa_y \langle T^1_x \rangle \aleph) + i \langle O^1_z \rangle \Re |^2 \qquad (6.2)$$
$$+ | \sqrt{(3/2)} \langle T^1_z \rangle \aleph - i\kappa_x \langle O^1_y \rangle \aleph - \kappa_y \langle O^1_x \rangle \Re |^2 \}.$$

Our estimates of $\aleph$ and $\Re$ are listed in Table 2 for a few values of h and k. These estimates imply that $\aleph$ and $\Re$ as functions of h and k can vary by large factors, e.g., for the pair (11, 9, 0)$_m$ and (9, 11, 0)$_m$ values of $\aleph'$ and $\Re'$ differ by factors of 40.5 and 52.0, respectively, while we find $\aleph' \approx \Re' \approx 0$ and $\aleph'' \approx \Re'' \approx -1.0$ for (11, 19, 0)$_m$.

Recall that axes (x, y, z) and ($\mathbf{a}^*_m$, $\mathbf{b}_m$, $\mathbf{c}_m$) are chosen to coincide, and our monoclinic cell has unique $b_m$-axis. The magnetic dipole moment is reported to be parallel to the a-axis of the parent, orthorhombic structure [1]. If this experimental finding is correct $\langle T^1_x \rangle = \langle T^1_z \rangle = 0$ and $\langle T^1_y \rangle$ is different from zero, leaving an intensity,

$$| \langle \mathbf{Q}_\perp \rangle^{(+)} + \langle \mathbf{Q}_\perp \rangle^{(-)} |^2 \approx 96 | \Re |^2 \{ | \sqrt{(3/2)} \kappa_x \langle T^1_y \rangle + i \langle O^1_z \rangle |^2 + | \kappa_x \langle O^1_y \rangle (\aleph/\Re) - i\kappa_y \langle O^1_x \rangle |^2 \}. \quad (6.3)$$

This expression for the intensity of Bragg spots indexed by (h, k, 0)$_m$, with $\kappa_x \propto h$ and $\kappa_y \propto k$, can be tested against experimental data to unearth anapoles $\langle \mathbf{O}^1 \rangle$. In the event that (6.3) proves to be inadequate for this purpose the assumption $\langle T^1_x \rangle = \langle T^1_z \rangle = 0$ should be rescinded and data confronted with (6.2).

7. Resonant x-ray Bragg diffraction

Magnetic charge $\langle \mathbf{S} \cdot \mathbf{n} \rangle$ contributes to Bragg spots enhanced by an electric dipole - magnetic dipole (E1-M1) resonant event, in addition to an anapole and a quadrupole. Expressions for E1-M1 amplitudes in rotated and un-rotated channels of polarization expressed in terms of $\Psi^K_{\pm Q} = \mathcal{A}^K_Q \pm \mathcal{B}^K_Q$ are reported by Lovesey and Scagnoli [20], and we do not repeat them here. The resonant event is weak but nonetheless it can contribute in the soft x-ray region [5]. Expressions for $\mathcal{A}^K_Q$ and $\mathcal{B}^K_Q$ are derived from (4.2) for BaFe$_2$Se$_3$. The Bragg spot $(1, 1, 0)_m$ yields a Bragg angle θ determined by sin θ = (0.0576 Å$^{-1}$) λ, with an x-ray wavelength λ = (12.4/E) Å and E in units of keV. Using E = 0.721 keV for the Fe L$_2$-edge yields θ = 82.5°.

The parity-odd resonant event using electric dipole - electric quadrupole (E1-E2) operators reveals multipoles with ranks K = 1, 2 and 3. Amplitudes in terms of $\mathcal{A}^K_Q$ and $\mathcal{B}^K_Q$ are listed by Scagnoli and Lovesey [6], together with corresponding expressions for E1-E1 and E2-E2 amplitudes in terms of $A^K_Q$ and $B^K_Q$.

8. Discussion

Based on data for BaFe$_2$Se$_3$ reported by Caron J M et al [1] we submit that long-range magnetic order below $T_N$ = 256 K belongs to the space-group C$_a$c (#9.41) with polar crystal-class m1' [2, 16, 17]. Chemical and magnetic structures are depicted in Figure1. Ferrous ions occupy sites that have no symmetry (site symmetry 1). Magneto-electric multipoles that are both time-odd and parity-odd are allowed in BaFe$_2$Se$_3$, as a consequence of the absence of a centre of inversion symmetry in the site symmetry together with long-range magnetic order. Of these multipoles perhaps the best known are a magnetic charge (monopole) and an anapole (toroidal dipole moment).

Magneto-electric multipoles are constructed from electronic states with different orbital angular momenta. In this respect they have the same property as the (local) electric dipole moment, which is a polar vector but time-even. Orbital anapoles and electric dipole moments usually point in different directions in a crystal, with orbital anapoles having a larger magnitude. Electronic states that contribute to a non-zero magneto-electric multipole can be inter-ion, e.g., 3d(Fe) - 4p(Fe), and admixtures of 3d(Fe) and p-type ligand orbitals that participate in transport properties and bonding.

In view of the important electronic information stored in magneto-electric multipoles it is fortunate that they contribute to intensities of Bragg spots created in the diffraction of neutrons and x-rays. Our simulation of Bragg diffraction has been tackled with a formulation that can be applied with equal ease to neutron diffraction and x-ray diffraction, be it Thomson (Templeton and Templeton scattering) or scattering enhanced by an atomic resonance. Expressions for resonant x-ray diffraction exist in the literature [6, 20] and this communication reports the counterpart expression for neutron diffraction (published expressions for resonant x-ray diffraction include all multipoles allowed in the four channels of polarization, while our expressions for neutron diffraction reported in an Appendix are cut short at a level of multipoles that are likely responsible for the dominant contributions to the intensities of neutron Bragg spots).

A low symmetry of the BaFe$_2$Se$_3$ magnetic structure means few selection rules exist to eliminate some multipoles from the unit-cell structure factor. One manifestation of the low symmetry is that magnetic dipoles and anapoles are not constrained to lie along particular crystal directions. By way of illustration of our general result we explore intensities indexed by (h, k, 0)$_m$, where Miller indices h and k are integers, in terms of magnetic dipoles and anapoles. Intensity of a Bragg spot contains six dipoles, because magnetic dipoles and anapoles are not constrained by symmetry. Existing data suggests that magnetic dipoles actually point along the a-axis of the parent structure, and with this information observed intensities can be parameterized in terms of three quantities that are the ratios of the three components of the anapole to the magnetic moment.

**Acknowledgement** One of us (SWL) is grateful to Professor E Balcar for helpful correspondence.

**Appendix**

An expression for $\langle \mathbf{Q} \rangle$ is derived using $\Psi^K_{\pm Q} = A^K_Q \pm B^K_Q$ for parity-even multipoles in magnetic neutron diffraction [12, 25, 26], i.e., $\langle U^K_Q \rangle = \langle T^K_Q \rangle$ in (4.1), and we add a superscript (+) to $\langle \mathbf{Q} \rangle$ to denote the parity signature $\sigma_\pi = +1$. It is universal and not explicit for the material of interest in this communication. A unit vector $\boldsymbol{\kappa} = \mathbf{k}/k$ defines the direction of the Bragg wavevector in orthogonal coordinates (x, y, z). With the chosen notation $\Psi^1_x = -\sqrt{2} B^1_1$, $\Psi^1_y = i\sqrt{2} A^1_1$ and $\Psi^1_z = A^1_0$. Retaining dipoles (K = 1), quadrupoles (K = 2) and octupoles (K = 3) in the general expression [12, 25] we find,

$\langle \mathbf{Q}_x \rangle^{(+)} \approx -(3/\sqrt{2}) B^1_1 + \sqrt{3}\{\kappa_y \kappa_z [A^2_2 + \sqrt{(3/2)} A^2_0] + \kappa_x [i\kappa_z B^2_2 - \kappa_y B^2_1] + i(\kappa_y^2 - \kappa_z^2) A^2_1\}$

$+ (3/4)\sqrt{35}\{\kappa_x \kappa_z [\sqrt{(2/3)} A^3_2 - \sqrt{(1/5)} A^3_0] - \sqrt{(1/15)} (3\kappa_z^2 - 1) B^3_1$    (A1)

$+ i\kappa_x \kappa_y [A^3_3 - \sqrt{(1/15)} A^3_1] - i\sqrt{(2/3)} \kappa_y \kappa_z B^3_2 + (1/2) (\kappa_x^2 - \kappa_y^2) [\sqrt{(1/15)} B^3_1 - B^3_3]\}$,

$\langle \mathbf{Q}_y \rangle^{(+)} \approx (3i/\sqrt{2}) A^1_1 + \sqrt{3}\{\kappa_x \kappa_z [A^2_2 - \sqrt{(3/2)} A^2_0] - i\kappa_y [\kappa_x A^2_1 + \kappa_z B^2_2] + (\kappa_x^2 - \kappa_z^2) B^2_1\}$

$+ (3/4)\sqrt{35}\{-\kappa_y \kappa_z [\sqrt{(2/3)} A^3_2 + \sqrt{(1/5)} A^3_0] + i\sqrt{(1/15)} (3\kappa_z^2 - 1) A^3_1$    (A2)

$+ \kappa_x \kappa_y [B^3_3 + \sqrt{(1/15)} B^3_1] - i\sqrt{(2/3)} \kappa_x \kappa_z B^3_2 + (i/2) (\kappa_x^2 - \kappa_y^2) [\sqrt{(1/15)} A^3_1 + A^3_3]\}$,

$\langle \mathbf{Q}_z \rangle^{(+)} \approx (3/2) A^1_0 + \sqrt{3}\{\kappa_z [i\kappa_x A^2_1 + \kappa_y B^2_1] - 2\kappa_x \kappa_y A^2_2 - i(\kappa_x^2 - \kappa_y^2) B^2_2\}$    (A3)

$+ (1/4)\sqrt{7}\{(3/2) (3\kappa_z^2 - 1) A^3_0 + 4\sqrt{3} \kappa_z [i\kappa_y A^3_1 - \kappa_x B^3_1] + \sqrt{(15/2)} [(\kappa_x^2 - \kappa_y^2) A^3_2 - 2i\kappa_x \kappa_y B^3_2]\}$.

The scattering amplitude for magnetic neutron Bragg diffraction $\langle \mathbf{Q}_\perp \rangle$ can be constructed using $\langle \mathbf{Q}_\perp \rangle = [\langle \mathbf{Q} \rangle - \boldsymbol{\kappa} (\boldsymbol{\kappa} \cdot \langle \mathbf{Q} \rangle)]$.

Universal expressions for Cartesian components of the intermediate amplitude for parity-odd multipoles, $\langle \mathbf{Q} \rangle^{(-)}$ [14], at the chosen level of approximation, are simpler expressions than we encountered above for parity-even multipoles, because now we encounter quadrupoles rather than octupoles, in addition to dipoles (anapoles). The relative simplicity of $\langle \mathbf{Q} \rangle^{(-)}$ prompts us to construct explicit expressions for the amplitude $\langle \mathbf{Q}_\perp \rangle^{(-)}$. For parity-odd multipoles we adopt a notation $\Psi^K_{\pm Q} = \mathcal{A}^K_Q \pm \mathcal{B}^K_Q$, with $\mathcal{A}^1_Q$ and $\mathcal{B}^1_Q$ composed from $\langle \mathbf{H}^1 \rangle$ and $\langle \mathbf{O}^1 \rangle$, and $\mathcal{A}^2_Q$ and $\mathcal{B}^2_Q$ composed from quadrupoles $\langle \mathbf{H}^2_Q \rangle$ only [14]. We go on to find,

$\langle \mathbf{Q}_{\perp,x} \rangle^{(-)} \approx -\sqrt{3}[(i/\sqrt{2}) \kappa_y \mathcal{A}^1_0 + \kappa_z \mathcal{A}^1_1] + (3/\sqrt{5}) [i\kappa_x\{(2\kappa_y^2 + \kappa_z^2) \mathcal{A}^2_2 - \sqrt{(3/2)} \kappa_z^2 \mathcal{A}^2_0\}$

$+ (1 - 2\kappa_x^2)(\kappa_y \mathcal{B}^2_2 - i\kappa_z \mathcal{B}^2_1) + 2\kappa_x \kappa_y \kappa_z \mathcal{A}^2_1]$,    (A4)

$$\langle \mathbf{Q}_{\perp,y}\rangle^{(-)} \approx i\sqrt{3}[(1/\sqrt{2})\ \kappa_x\ \mathcal{A}^1_0 + \kappa_z\ \mathcal{B}^1_1] + (3/\sqrt{5})\ [-i\kappa_y\{(2\kappa_x^2 + \kappa_z^2)\ \mathcal{A}^2_2 + \sqrt{(3/2)}\ \kappa_z^2\ \mathcal{A}^2_0\}$$

$$+ (1 - 2\kappa_y^2)(\kappa_x\ \mathcal{B}^2_2 - \kappa_z\ \mathcal{A}^2_1) + 2i\ \kappa_x\kappa_y\kappa_z\ \mathcal{B}^2_1], \qquad (A5)$$

$$\langle \mathbf{Q}_{\perp,z}\rangle^{(-)} \approx \sqrt{3}[\kappa_x\ \mathcal{A}^1_1 - i\kappa_y\ \mathcal{B}^1_1] + (3/\sqrt{5})\ [i\sqrt{(3/2)}\ \kappa_z\ (\kappa_x^2 + \kappa_y^2)\ \mathcal{A}^2_0 - i\kappa_z\ (\kappa_x^2 - \kappa_y^2)\ \mathcal{A}^2_2$$

$$- (1 - 2\kappa_z^2)(\kappa_y\ \mathcal{A}^2_1 + i\kappa_x\ \mathcal{B}^2_1) - 2\ \kappa_x\kappa_y\kappa_z\ \mathcal{B}^2_2]. \qquad (A6)$$

Higher-order multipoles omitted from (A.1) - (A.6) can be found in references [12, 14, 25, 26].

---

Table 1. Estimated structure parameters for the magnetic space-group $C_ac$ (#9.41) derived from the orthorhombic parent structure Pnma [1].

| j | $x_j$ | $y_j$ | $z_j$ |
| --- | --- | --- | --- |
| 1 | 0.6995 | 0.1215 | 0.3970 |
| 2 | 0.0505 | 0.3715 | 0.6030 |
| 3 | 0.3025 | 0.1285 | 0.1030 |
| 4 | 0.4475 | 0.3785 | 0.8970 |

---

Table 2. Typical values of $\aleph = \aleph' + i\aleph''$ and $\Re = \Re' + i\Re''$ that appear in equation (6.2) for a dipole approximation to the intensity of Bragg spots $(h, k, 0)_m$ with both h and k odd. They are derived from $\aleph = \sum_j f_j \operatorname{Re}(g_j)$ and $\Re = \sum_j f_j \operatorname{Im}(g_j)$ using the definitions $f_j = \exp(2\pi i(x_j h + z_j l))$ and $g_j = \exp(2\pi i y_j k)$, with $l = 0$ and structure parameters listed in Table 1.

| h | k | $\aleph'$ | $\aleph''$ | $\Re'$ | $\Re''$ |
|---|---|---|---|---|---|
| 1 | 1 | −0.4227 | −0.4821 | −0.4174 | 0.4760 |
| 5 | 1 | −1.4575 | −1.4545 | −1.3687 | 1.3658 |
| 1 | 3 | 0.5405 | 0.3626 | −0.5337 | 0.3580 |
| 5 | 3 | 1.4486 | 1.4577 | −1.3604 | 1.3689 |
| 7 | 3 | −1.0927 | 0.9788 | 1.1933 | 1.0689 |
| 3 | 7 | 1.0113 | −1.2515 | −1.0502 | −1.2996 |
| 11 | 9 | −0.7585 | 0.2865 | −0.8713 | −0.3291 |
| 9 | 11 | −0.0188 | 0.6773 | 0.0167 | 0.6047 |
| 19 | 5 | 0.0311 | −0.2998 | 0.0395 | 0.3815 |
| 5 | 19 | 1.3032 | 1.3589 | −1.2238 | 1.2761 |
| 11 | 19 | 0.0 | −0.9740 | 0.0 | −1.1189 |

-----------------------------------------------------------------------------------------------------------------------

Figure 1. Chemical (right panel) and magnetic (left panel) structure of $BaFe_2Se_3$ based on the Pnma-type parent. Ladders of magnetic dipole moments run along the $a_m$-axis. Cell lengths a, b, c are depicted with a = 11.8834 Å, b = 5.4141 Å, c = 9.1409 Å [1]. Magnetic structure $C_ac$ (#9.41) is polar and belongs to the crystal class m1' [16, 17]. Basis vectors of the monoclinic cell in the magnetic space-group are {(0, 2, 0), (2, 0, 0), (0, −1, −1)} with respect to the parent and unique $b_m$-axis. Cell dimensions $a_m$ = 10.8282 Å, $b_m$ = 23.7668 Å, $c_m$ = 10.6240 Å and obtuse β = 120.638°. Estimated structure parameters are listed in Table 1.

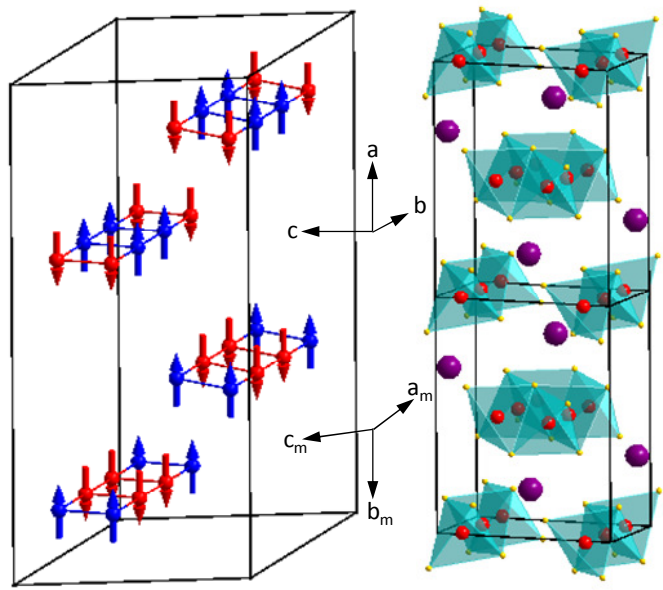